\documentclass[10pt]{article}

\RequirePackage[top=0.6in,left=0.6in,right=0.6in,bottom=0.3in,includefoot]{geometry}

\usepackage{graphicx}
\usepackage{hyperref}
\usepackage{amssymb,amsmath}
\usepackage[utf8]{inputenc}
\usepackage{longtable}
\usepackage{booktabs}
\usepackage{float}
\usepackage{calc}
\usepackage{multirow}
\usepackage{authblk}  
\usepackage{flushend} 
\usepackage[misc,geometry]{ifsym} 

\usepackage[T1]{fontenc}
\usepackage{lmodern}  
\usepackage{microtype}
\SetTracking{encoding=*, family=tt*}{-50}  

\renewcommand{\texttt}[1]{{\footnotesize\ttfamily\textls[-50]{#1}}}

\usepackage{xcolor}
\definecolor{bluehaze}{RGB}{54,95,145}
\definecolor{bluetint}{RGB}{0,50,100}
\definecolor{graytint}{RGB}{175,175,175}
\definecolor{graybit}{RGB}{245,245,245}
\definecolor{lightblack}{RGB}{75,75,75}

\hypersetup{
  breaklinks=true,
  pdfborder={0 0 0},
  colorlinks=true,
  urlcolor=bluehaze,
  linkcolor=bluehaze,
  citecolor=bluehaze
}

\usepackage[skip=2pt, font=footnotesize, format=plain, labelfont=it, textfont=it]{caption}
\DeclareCaptionLabelFormat{nolabel}{}
\usepackage{titlesec}
\titleformat{\section}
  {\normalfont\fontfamily{phv}\fontsize{11pt}{11pt}
    \selectfont\color{bluetint}\bfseries}{\thesection}{1em}{}
\titleformat{\subsection}
  {\normalfont\fontfamily{phv}\fontsize{9pt}{9pt}
    \selectfont\color{black}\bfseries}{\thesubsection}{1em}{}
\titleformat{\subsubsection}
  {\normalfont\fontfamily{phv}\fontsize{9pt}{9pt}
    \selectfont\color{black}}{\thesubsubsection}{1em}{}

\titlespacing\section{0pt}{12pt plus 2pt minus 2pt}{-2pt plus 2pt minus 2pt}
\titlespacing\subsection{0pt}{11pt plus 2pt minus 2pt}{-2pt plus 2pt minus 2pt}
\titlespacing\subsubsection{0pt}{4pt plus 2pt minus 2pt}{-2pt plus 2pt minus 2pt}

\usepackage{needspace}
\let\oldsection\section
\renewcommand{\section}[1]{\needspace{4\baselineskip}\oldsection{#1}}

\usepackage{enumitem}
\setlist[itemize]{topsep=0pt,parsep=2pt,leftmargin=18pt}

\usepackage{mdframed}
\newmdenv[linecolor=white, fontcolor=bluetint, backgroundcolor=graybit,
  leftmargin=8, rightmargin=8, innertopmargin=8, innerbottommargin=8,
  innerleftmargin=12, innerrightmargin=12, roundcorner=10pt,
  font=\normalsize]{infobox}

\usepackage{fancyhdr}
\usepackage{color}
\usepackage{fancyvrb}

\DefineVerbatimEnvironment{Highlighting}{Verbatim}{commandchars=\\\{\}}
\newenvironment{Shaded}{}{}

\newcommand{\DecValTok}[1]{\textcolor[rgb]{0.25,0.63,0.44}{#1}}

\newcommand{\ImportTok}[1]{#1}

\newcommand{\NormalTok}[1]{#1}
\newcommand{\OperatorTok}[1]{\textcolor[rgb]{0.40,0.40,0.40}{#1}}

\newcommand{\StringTok}[1]{\textcolor[rgb]{0.25,0.44,0.63}{#1}}

\makeatletter
\renewenvironment{Shaded}{\footnotesize}{}
\makeatother

\newlength{\cslhangindent}
\newlength{\csllabelwidth}
\newenvironment{CSLReferences}[2]
  {\setlength{\parindent}{0pt}%
   \everypar{\setlength{\hangindent}{\cslhangindent}}\ignorespaces}
  {\par}

\newcommand{\CSLLeftMargin}[1]{\parbox[t]{\csllabelwidth}{#1}}
\newcommand{\CSLRightInline}[1]{\parbox[t]{\linewidth - \csllabelwidth}{#1}\break}

      \author[1,2]{Nathan J. LeRoy}
        \author[1]{Donald R Campbell Jr}
        \author[3]{Seth Stadick}
        \author[1]{Oleksandr Khoroshevskyi}
        \author[1]{Sang-Hoon Park}
        \author[1]{Ziyang Hu}
        \author[1,2,4,5,\Letter]{Nathan C. Sheffield}
  
\affil[1]{Department of Genome Sciences, School of Medicine, University
of Virginia, 22908, Charlottesville VA}
\affil[2]{Department of Biomedical Engineering, School of Medicine,
University of Virginia, 22908, Charlottesville VA}
\affil[3]{Life Sciences Group, Bio-Rad Laboratories, 1000 Alfred Nobel
Dr, Hercules, 94547, California, USA}
\affil[4]{Department of Biochemistry and Molecular Genetics, School of
Medicine, University of Virginia, 22908, Charlottesville VA}
\affil[5]{School of Data Science, University of Virginia, 22908,
Charlottesville, VA}

                  \affil[ ]{\Letter \hspace{0.08cm} Correspondence: \href{mailto:nsheffield@virginia.edu}{nsheffield@virginia.edu}}

\title{Fast, memory-efficient genomic interval tokenizers for modern
machine learning}

\usepackage{titling}

\makeatletter
\renewcommand{\maketitle}{
  \begin{raggedright}
  \huge\textbf{\@title}
    \par\normalsize \vspace{0.5em}
    {    \ignorespaces\@author
        \par}
    \vspace{0.2em}
  \end{raggedright}
}
\makeatother

\begin{document}

    \twocolumn[{%
      \begin{@twocolumnfalse}
                \maketitle
        
                \begin{infobox}
        \textbf{Introduction:} Epigenomic datasets from high-throughput
        sequencing experiments are commonly summarized as genomic
        intervals. As the volume of this data grows, so does interest in
        analyzing it through deep learning. However, the heterogeneity
        of genomic interval data, where each dataset defines its own
        regions, creates barriers for machine learning methods that
        require consistent, discrete vocabularies. \textbf{Methods:} We
        introduce gtars-tokenizers, a high-performance library that maps
        genomic intervals to a predefined universe or vocabulary of
        regions, analogous to text tokenization in natural language
        processing. Built in Rust with bindings for Python, R, CLI, and
        WebAssembly, gtars-tokenizers implements two overlap methods
        (BITS and AIList) and integrates seamlessly with modern ML
        frameworks through Hugging Face-compatible APIs.
        \textbf{Results:} The gtars-tokenizers package achieves top
        efficiency for large-scale datasets, while enabling genomic
        intervals to be processed using standard ML workflows in PyTorch
        and TensorFlow without ad hoc preprocessing. This token-based
        approach bridges genomics and machine learning, supporting
        scalable and standardized analysis of interval data across
        diverse computational environments. \textbf{Availability:} PyPI
        and GitHub: \url{https://github.com/databio/gtars}.
        \end{infobox}
        
              \end{@twocolumnfalse}
    }]

\hypertarget{introduction}{%
\section{Introduction}\label{introduction}}

Advancements in high-throughput sequencing technologies have generated
vast and diverse epigenomic datasets from assays such as ChIP-seq,
ATAC-seq, and Hi-C\textsuperscript{1,2}. These experiments are
frequently summarized as genomic intervals, which define regions on a
genome. Summarized genomic interval data has grown rapidly over the past
few years\textsuperscript{3,4}. This proliferation of data provides a
valuable opportunity to uncover generalizable patterns, support
predictive modeling, and enable transfer learning using large-scale
machine learning (ML) methods. However, a major barrier in applying
modern ML methods to genomic interval data arises due to the
heterogeneity of the data. Genomic interval data is inherently variable
and unstructured; each dataset defines its own regions of interest,
making it difficult to compare or combine results across experiments.
This is incompatible with ML methods, which generally require data to be
described in a discrete, consistent vocabulary. For example, in natural
language processing (NLP), models require well-defined vocabularies to
process and integrate diverse sets of textual data. The process of
mapping new, unseen datasets to a shared feature set is called
\emph{tokenization} and is a vital part of NLP research and
development\textsuperscript{5--8}. Without such a standardized basis, it
is difficult or impossible to create feature-aligned representations
suitable for ML. Similarly, for genomic intervals, it is necessary to
map new datasets to a shared vocabulary, or \emph{consensus} set of
genomic intervals\textsuperscript{4,9}. This process is conceptually
similar to tokenization in NLP. It serves the same purpose: to enable
consistent and scalable representation of variable input data.

The basis of mapping genomic intervals into a shared, consensus set lies
in general-purpose interval comparison. While many tools exist for
genomic interval comparison\textsuperscript{10--15}, they are limited in
several ways. First, they are typically only accessible in a single
environment, such as R, or as command-line tools. They are not optimized
for fast, in-memory processing. This limitation poses a significant pain
point for machine learning pipelines in Python, which require
high-throughput, efficient data handling. Second, they generally lack
flexible APIs that integrate seamlessly in the Python-based machine
learning ecosystem, particularly with libraries like \texttt{PyTorch},
\texttt{TensorFlow}, or \texttt{huggingface/transformers.} As a result,
ML applications in genomics often suffer from ad hoc preprocessing
steps, pipeline bottlenecks, and limited scalability.

In our recent work to develop Atacformer, a transformer-based foundation
model for ATAC-seq data, we wanted to streamline the process of
tokenizing genomic intervals for deep learning\textsuperscript{16}. To
address this, we created \texttt{gtars-tokenizers}, a library designed
specifically for genomic machine learning. It provides four main
advantages over existing tools: First, the Rust core is faster than many
existing tools and rivals the fastest current implementations. Second,
it exposes a direct bridge into HuggingFace and PyTorch so genomic
intervals can be used without ad hoc preprocessing, a convenience for
modern ML infrastructure. Third, it treats genomic intervals in a way
that mirrors the conceptualization of words in NLP, enabling consistent,
vocabulary-based representations that scale across datasets. Finally, it
offers a unified engine with bindings for Python, R, Rust, command line,
and web applications so the same foundation can serve diverse users and
workflows. Together, these advances make \texttt{gtars-tokenizers} a
valuable step for machine learning on genomic interval data.

\setlength{\intextsep}{2pt}\setlength{\columnsep}{8pt}\begin{figure*}[t]\centering\includegraphics{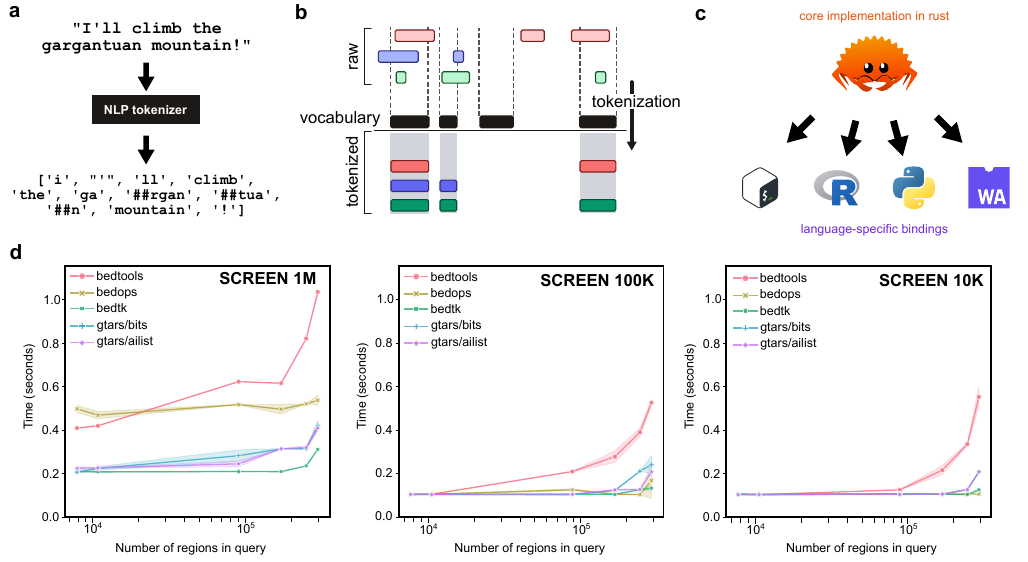}\caption{\textbf{\label{overview}Figure
\ref{overview}. Overview and benchmarking of gtokenizers, a Rust-based
library for genomic interval tokenization.} \textbf{a,} Schematic of
natural language tokenization. NLP tokenizers typically break sentences
up into words or word-pieces. \textbf{b,} Schematic illustrating
gtokenizers applied to regulatory elements (e.g., cCREs) for
standardized interval representation. \textbf{c,} Architecture of
gtokenizers, with a core implementation in Rust and support for multiple
language bindings (e.g., CLI, R, Python, WebAssembly). \textbf{d,}
Runtime benchmarking across three query sizes (1M, 100K, 10K regions)
against existing tools (bedtools, bedops, bedtk) and Rust-based
implementations (\texttt{gtars/bits}, \texttt{gtars/alist}),
demonstrating scalability and performance.}\vspace{-5pt}\end{figure*}

\hypertarget{results}{%
\section{Results}\label{results}}

\hypertarget{overview-of-the-genomic-interval-tokenizers}{%
\subsection{Overview of the genomic interval
tokenizers}\label{overview-of-the-genomic-interval-tokenizers}}

Modern deep-learning workflows in NLP require tokenizers to convert new
text into the model's fixed vocabulary, enabling consistent inputs for
downstream processing. Tokens in language models correspond to discrete
words or subword units (Fig. \ref{overview}A). In genomics, a comparable
process is necessary: machine learning models that treat genomic
intervals as discrete units, like words in a sentence, must map each
dataset to a common set of regions, or a vocabulary for genomic
intervals\textsuperscript{9,17--20}. This vocabulary ensures data across
experiments are represented in a standardized, comparable way (Fig.
\ref{overview}B). Different datasets can thus be interpreted with the
same model architecture and feature space, just as diverse text inputs
are aligned via tokenization in NLP.

We implemented two overlap methods in \texttt{gtars-tokenizers}:
\texttt{gtars/bits}, which uses binary interval tree search
(BITS)\textsuperscript{21}, and \texttt{gtars/alist}, which uses an
Augmented Interval List (AIList)\textsuperscript{13}. Both methods are
implemented in Rust for performance and memory efficiency. To maximize
flexibility and usability, we provide bindings for
\texttt{gtars/tokenizers} in Python, R, and WebAssembly, as well as a
command-line interface (CLI) (Fig. \ref{overview}C). This allows users
to integrate genomic interval tokenization into their existing
workflows, whether they are using Python-based machine learning
libraries like TensorFlow or PyTorch, R-based bioinformatics tools, or
require a web-based solution for use in a browser.

\hypertarget{gtars-tokenizers-are-highly-performant}{%
\subsection{Gtars tokenizers are highly
performant}\label{gtars-tokenizers-are-highly-performant}}

To highlight the performance of \texttt{gtars/tokenizers}, we
benchmarked it against existing tools for genomic interval tokenization.
We compare \texttt{gtars/tokenizers} to \texttt{bedtools},
\texttt{bedops}, and \texttt{bedtk}. These tools focus on
general-purpose genomic interval arithmetic and are not optimized for
machine learning applications. We found \texttt{gtars/tokenizers} to be
consistently \emph{as fast as or faster than} existing tools (Fig.
\ref{overview}D). For large universes with \textgreater1 million
intervals (like those used in genomic interval machine learning),
\texttt{gtars-tokenizers} is around 2- 3x faster than \texttt{bedtools}
and \texttt{bedops}, while being comparable to \texttt{bedtk}. This
pattern holds across different query sizes (1M, 100K, and 10K regions),
demonstrating the scalability and performance of
\texttt{gtars/tokenizers}.

\hypertarget{section}{%
\subsection{}\label{section}}

\hypertarget{gtars-works-seamlessly-with-modern-machine-learning-infrastructure}{%
\subsection{Gtars works seamlessly with modern machine learning
infrastructure}\label{gtars-works-seamlessly-with-modern-machine-learning-infrastructure}}

The \texttt{gtars-tokenizer} implementation is compatible with the
Hugging Face tokenizers API, enabling seamless integration with the
broader Hugging Face ecosystem. The \texttt{gtars} tokenizers are
near-drop-in replacements for existing Hugging Face tokenizers, meaning
users can pass them to the HuggingFace \texttt{transformers} package
functions and classes using the same ergonomics as a standard NLP
workflow. The consistent interface makes it easy for ML engineers to
adapt to training models on genomic interval data. It also means that
the downstream outputs of the training process will seamlessly integrate
with popular downstream frameworks and tools that rely on the Hugging
Face tokenizers standard, such as PyTorch Lightning, AllenNLP, and
evaluation libraries like Evaluate, PEFT, and Weights \& Biases. For
example, this is how one can use our tokenizers to preprocess data for a
simple neural network built with PyTorch by first creating a new
tokenizer from a BED-file, and then preprocessing data for a neural
network:

\begin{Shaded}
\begin{Highlighting}[]
\ImportTok{import}\NormalTok{ torch  }
\ImportTok{import}\NormalTok{ gtars.tokenizers }\ImportTok{as}\NormalTok{ Tokenizer}

\NormalTok{tokenizer }\OperatorTok{=}\NormalTok{ Tokenizer.from\_bed(}\StringTok{"path/to/file.bed"}\NormalTok{)  }
\NormalTok{network }\OperatorTok{=}\NormalTok{ torch.nn.Embedding(tokenizer.vocab\_size, }\DecValTok{64}\NormalTok{)}

\NormalTok{query\_intervals }\OperatorTok{=}\NormalTok{ [(}\StringTok{"chr1"}\NormalTok{, }\DecValTok{100}\NormalTok{, }\DecValTok{200}\NormalTok{), (}\StringTok{"chr2"}\NormalTok{, }\DecValTok{300}\NormalTok{, }\DecValTok{400}\NormalTok{)]  }
\NormalTok{tokens }\OperatorTok{=}\NormalTok{ tokenizer.tokenize(query\_intervals)[}\StringTok{"input\_ids"}\NormalTok{]  }
\NormalTok{out }\OperatorTok{=}\NormalTok{ network(torch.tensor(input\_ids))  }
\end{Highlighting}
\end{Shaded}

\hypertarget{gtars-tokenizers-are-available-in-a-wide-array-of-computing-environments}{%
\subsection{Gtars tokenizers are available in a wide array of computing
environments}\label{gtars-tokenizers-are-available-in-a-wide-array-of-computing-environments}}

To maximize usability, we expose the Rust core of
\texttt{gtars-tokenizers} as a Rust library crate, as a command-line
tool, with R bindings, Python bindings, and for WebAssembly (WASM). This
broad set of interfaces ensures that the same high-performance engine
can serve diverse communities, including machine learning researchers,
bioinformaticians, and end-users in web tools, without duplicating
functionality or compromising performance. It also reduces maintenance
requirements because a single fast interface can be deployed in many
situations.

\hypertarget{discussion}{%
\section{Discussion}\label{discussion}}

The \texttt{gtars-tokenizers} project provides a valuable tool for ML in
genomics. It facilitates our work building ML models for genomic
intervals\textsuperscript{16}, and will also have many broader uses. A
fast, unified interface makes it easy to integrate into the existing
popular ML packages while also being useful for traditional applications
of interval overlap arithmetic. New tools like \texttt{gtars} will be an
important part of the evolving ecosystem that promote fast analysis on
genomic regions. Future work that will extend this approach to
fragments, AnnData objects, bulk ATAC-seq, or even SNPs could reshape
how we represent and analyze the genome. By providing a fast, ML-aware
abstraction, \texttt{gtars-tokenizers} moves the field beyond
tool-specific pipelines toward interoperable, general-purpose models of
genome function.

\hypertarget{funding-statement}{%
\subsection{Funding statement}\label{funding-statement}}

This work was supported by National Human Genome Research Institute
grant R01-HG012558 (NCS).

\hypertarget{conflict-of-interest-statement}{%
\subsection{Conflict of interest
statement}\label{conflict-of-interest-statement}}

NCS is a consultant for InVitro Cell Research, LLC. All other authors
report no conflicts of interest.

\hypertarget{references}{%
\section{References}\label{references}}

\hypertarget{refs}{}
\begin{CSLReferences}{0}{0}
\leavevmode\vadjust pre{\hypertarget{ref-Sheffield2012}{}}%
\CSLLeftMargin{1. }
\CSLRightInline{Sheffield, N. C. \& Furey, T. S. Identifying and
characterizing regulatory sequences in the human genome with chromatin
accessibility assays. \emph{Genes} \textbf{3,} 651--70 (2012).}

\leavevmode\vadjust pre{\hypertarget{ref-Lee2023}{}}%
\CSLLeftMargin{2. }
\CSLRightInline{Lee, J.-Y. The principles and applications of
high-throughput sequencing technologies. \emph{Development \&amp;
Reproduction} \textbf{27,} 9--24 (2023).}

\leavevmode\vadjust pre{\hypertarget{ref-Khoroshevskyi2023}{}}%
\CSLLeftMargin{3. }
\CSLRightInline{Khoroshevskyi, O., LeRoy, N. J., Reuter, V. P. \&
Sheffield, N. C. GEOfetch: A command-line tool for downloading data and
standardized metadata from GEO and SRA. \emph{Bioinformatics} (2023).
doi:\href{https://doi.org/10.1093/bioinformatics/btad069}{10.1093/bioinformatics/btad069}}

\leavevmode\vadjust pre{\hypertarget{ref-Xue2023}{}}%
\CSLLeftMargin{4. }
\CSLRightInline{Xue, B., Khoroshevskyi, O., Gomez, R. A. \& Sheffield,
N. C. Opportunities and challenges in sharing and reusing genomic
interval data. \emph{Frontiers in Genetics} \textbf{14,} (2023).}

\leavevmode\vadjust pre{\hypertarget{ref-Sennrich2016}{}}%
\CSLLeftMargin{5. }
\CSLRightInline{Sennrich, R., Haddow, B. \& Birch, A. Neural {Machine
Translation} of {Rare Words} with {Subword Units}. (2016).
doi:\href{https://doi.org/10.48550/arXiv.1508.07909}{10.48550/arXiv.1508.07909}}

\leavevmode\vadjust pre{\hypertarget{ref-Wu2016}{}}%
\CSLLeftMargin{6. }
\CSLRightInline{Wu, J. \emph{et al.} The landscape of accessible
chromatin in mammalian preimplantation embryos. \emph{Nature} (2016).
doi:\href{https://doi.org/10.1038/nature18606}{10.1038/nature18606}}

\leavevmode\vadjust pre{\hypertarget{ref-Kudo2018}{}}%
\CSLLeftMargin{7. }
\CSLRightInline{Kudo, T. Subword {Regularization}: {Improving Neural
Network Translation Models} with {Multiple Subword Candidates}. (2018).
doi:\href{https://doi.org/10.48550/arXiv.1804.10959}{10.48550/arXiv.1804.10959}}

\leavevmode\vadjust pre{\hypertarget{ref-Kudo2018a}{}}%
\CSLLeftMargin{8. }
\CSLRightInline{Kudo, T. \& Richardson, J. {SentencePiece}: {A} simple
and language independent subword tokenizer and detokenizer for {Neural
Text Processing}. (2018).
doi:\href{https://doi.org/10.48550/arXiv.1808.06226}{10.48550/arXiv.1808.06226}}

\leavevmode\vadjust pre{\hypertarget{ref-Rymuza2024}{}}%
\CSLLeftMargin{9. }
\CSLRightInline{Rymuza, J. \emph{et al.} Methods for constructing and
evaluating consensus genomic interval sets. \emph{Nucleic Acids
Research} (2024).
doi:\href{https://doi.org/10.1093/nar/gkae685}{10.1093/nar/gkae685}}

\leavevmode\vadjust pre{\hypertarget{ref-Quinlan2010}{}}%
\CSLLeftMargin{10. }
\CSLRightInline{Quinlan, A. R. \& Hall, I. M. {BEDTools}: A flexible
suite of utilities for comparing genomic features. \emph{Bioinformatics}
\textbf{26,} 841--842 (2010).}

\leavevmode\vadjust pre{\hypertarget{ref-Neph2012}{}}%
\CSLLeftMargin{11. }
\CSLRightInline{Neph, S. \emph{et al.} An expansive human regulatory
lexicon encoded in transcription factor footprints. \emph{Nature}
\textbf{489,} 83--90 (2012).}

\leavevmode\vadjust pre{\hypertarget{ref-Li2021bedtk}{}}%
\CSLLeftMargin{12. }
\CSLRightInline{Li, H. \& Rong, J. Bedtk: Finding interval overlap with
implicit interval tree. \emph{Bioinformatics} \textbf{37,} 1315--1316
(2021).}

\leavevmode\vadjust pre{\hypertarget{ref-Feng2019}{}}%
\CSLLeftMargin{13. }
\CSLRightInline{Feng, J., Ratan, A. \& Sheffield, N. C. Augmented
interval list: A novel data structure for efficient genomic interval
search. \emph{Bioinformatics} (2019).
doi:\href{https://doi.org/10.1093/bioinformatics/btz407}{10.1093/bioinformatics/btz407}
PMCID:
\href{https://www.ncbi.nlm.nih.gov/pmc/articles/PMC6901075}{PMC6901075}}

\leavevmode\vadjust pre{\hypertarget{ref-Feng2021}{}}%
\CSLLeftMargin{14. }
\CSLRightInline{Feng, J. \& Sheffield, N. C. {IGD}: High-performance
search for large-scale genomic interval datasets. \emph{Bioinformatics}
\textbf{37,} 118--120 (2021). PMCID:
\href{https://www.ncbi.nlm.nih.gov/pmc/articles/33367484}{33367484}}

\leavevmode\vadjust pre{\hypertarget{ref-Schaefer2025}{}}%
\CSLLeftMargin{15. }
\CSLRightInline{Schäfer, R. A. \& Yang, R. A comprehensive benchmark of
tools for efficient genomic interval querying. \emph{Briefings in
Bioinformatics} \textbf{26,} (2025).}

\leavevmode\vadjust pre{\hypertarget{ref-LeRoy2025atacformer}{}}%
\CSLLeftMargin{16. }
\CSLRightInline{LeRoy, N. J. \emph{et al.} Atacformer: A
transformer-based foundation model for analysis and interpretation of
ATAC-seq data. \emph{bioRxiv} (2025).}

\leavevmode\vadjust pre{\hypertarget{ref-Gharavi2021}{}}%
\CSLLeftMargin{17. }
\CSLRightInline{Gharavi, E. \emph{et al.} Embeddings of genomic region
sets capture rich biological associations in low dimensions.
\emph{Bioinformatics} (2021).
doi:\href{https://doi.org/10.1093/bioinformatics/btab439}{10.1093/bioinformatics/btab439}
PMCID:
\href{https://www.ncbi.nlm.nih.gov/pmc/articles/PMC8652032}{PMC8652032}}

\leavevmode\vadjust pre{\hypertarget{ref-Gharavi2024}{}}%
\CSLLeftMargin{18. }
\CSLRightInline{Gharavi, E. \emph{et al.} Joint representation learning
for retrieval and annotation of genomic interval sets.
\emph{Bioengineering} \textbf{11,} 263 (2024).}

\leavevmode\vadjust pre{\hypertarget{ref-Zheng2024}{}}%
\CSLLeftMargin{19. }
\CSLRightInline{Zheng, G. L. \emph{et al.} Methods for evaluating
unsupervised vector representations of genomic regions. \emph{NAR
Genomics and Bioinformatics} \textbf{6,} (2024).}

\leavevmode\vadjust pre{\hypertarget{ref-LeRoy2024fast}{}}%
\CSLLeftMargin{20. }
\CSLRightInline{LeRoy, N. J. \emph{et al.} Fast clustering and cell-type
annotation of scATAC data using pre-trained embeddings. \emph{NAR
Genomics and Bioinformatics} \textbf{6,} (2024).}

\leavevmode\vadjust pre{\hypertarget{ref-Layer2013}{}}%
\CSLLeftMargin{21. }
\CSLRightInline{Layer, R. M., Skadron, K., Robins, G., Hall, I. M. \&
Quinlan, A. R. Binary interval search: A scalable algorithm for counting
interval intersections. \emph{Bioinformatics} \textbf{29,} 1--7 (2013).}

\end{CSLReferences}


\end{document}